\documentclass[lettersize,journal]{IEEEtran}
\usepackage{amsmath,amsfonts}
\usepackage{algorithmic}
\usepackage{algorithm}
\usepackage{array}
\usepackage[caption=false,font=normalsize,labelfont=sf,textfont=sf]{subfig}
\usepackage{textcomp}
\usepackage{stfloats}
\usepackage{url}
\usepackage{verbatim}
\usepackage{graphicx}
\usepackage{cite}
\usepackage{bm}
\begin{document}

\title{Design of a Gm-C Dynamic Amplifier with High Linearity and\\ High Temperature and Power Supply Voltage Stability}

\author{Jinkun\ Yang,\ Pengbin\ Xu
\thanks{The authors contributed equally to this work and should be considered co-first authors.}
\thanks{Jinkun Yang is with the School of Integrated Circuit Science and Engineering (Exemplary School of Microelectronics), University of Electronic Science and Technology of China, Chengdu, 611731, China (e-mail: 2022040906025@std.uestc.edu.cn).}
\thanks{Pengbin Xu is with the Institute of Integrated Circuits and Systems, University of Electronic Science and Technology of China, Chengdu, 611731, China (e-mail: 2231806072@qq.com).}}

\markboth{Manuscript, August~2025}%
{Shell \MakeLowercase{\textit{et al.}}: Design of a Gm-C Dynamic Amplifier with High Linearity and
 High Temperature and Power Supply Voltage Stability}

\IEEEpubid{}

\maketitle

\begin{abstract}
This paper presents a Gm-C dynamic amplifier with high linearity and high temperature and power supply voltage stability. The main part of the amplifier employs two asymmetric differential pairs to enhance transconductance linearity. The amplifier maintains a nearly constant gain within a differential input range of -40 mV to 40 mV, and achieves a total harmonic distortion (THD) of 70.5 dB. The bias part of the amplifier adopts a constant-gm bias circuit, which improves the temperature and supply voltage stability of the amplifier’s transconductance and gain. When the differential input is 1 mV, the power supply voltage fluctuates by $\bm{\pm}$10\%, and the temperature varies between -40$\bm{\mathrm{^\circ C}}$ and 120$\bm{\mathrm{^\circ C}}$, the standard deviation of the gain distribution is 262m, and the distribution range is from 15.1 to 16.3. 
\end{abstract}

\begin{IEEEkeywords}
ADC, Residue Amplifier, Gm-C dynamic amplifier, Linearization, Stability, THD.
\end{IEEEkeywords}

\section{Introduction}
\IEEEPARstart{T}{he} residue amplifier is a critical component in pipelined SAR ADCs. Errors introduced by the preceding ADC stage can be mitigated through inter-stage redundancy or by redundancy in the following ADC stage \cite{ref9}. However, errors introduced by the residue amplifier itself will be directly manifested in the digital output code of the following ADC stage. This requires the amplifier to have high linearity and stability against temperature and supply voltage variations. Open-loop Gm-C amplifiers, which achieve gain by charging a capacitor with a voltage-controlled current source of transconductance Gm, are promising for low-power, high-precision, or high-speed ADCs. Compared with closed-loop amplifiers, they inherently avoid stability issues \cite{ref5}, have a large time constant and narrow noise bandwidth that minimize noise contribution \cite{ref3}, and consume low power by operating only during the amplification phase \cite{ref4}. However, their performance is limited by high sensitivity to temperature and supply voltage variations, and significant linearity degradation under large input swings.

This paper presents a Gm-C dynamic amplifier with high linearity and high temperature and power supply voltage stability. While exerting the advantages of dynamic Gm-C amplifiers, such as low noise and low power consumption, it also compensates the limitations of traditional Gm-C amplifiers. Section II of this paper analyzes the principle of transconductance linearization, Section III outlines the structure of the proposed amplifier, Section IV presents the circuit layout and the simulation results, and Section V concludes the paper.

\section{The Principle of Transconductance Linearization}
To achieve linearization of the transconductance in a differential amplifier, the characteristic that the total transconductance of a differential pair shifts horizontally due to width-to-length ratio mismatch between input transistors\cite{ref1} is utilized.

\begin{figure}[htb]
\centering
\includegraphics[width=2.5in]{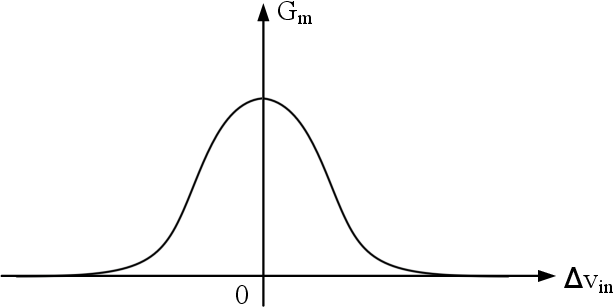}
\caption{$G_\mathrm{m}-\Delta V_\mathrm{in}$ curve of symmetric differential pair.}
\label{fig_1}
\end{figure}

In a symmetric differential pair, the curve of the total transconductance $G_\mathrm{m}$ versus the differential input voltage $\Delta V_\mathrm{{in}}$ is shown in Figure \ref{fig_1} \cite{ref1}. When $\Delta V_\mathrm{{in}}$=0, $G_\mathrm{m}$ reaches its maximum value.

\begin{figure}[htb]
\centering
\includegraphics[width=2.5in]{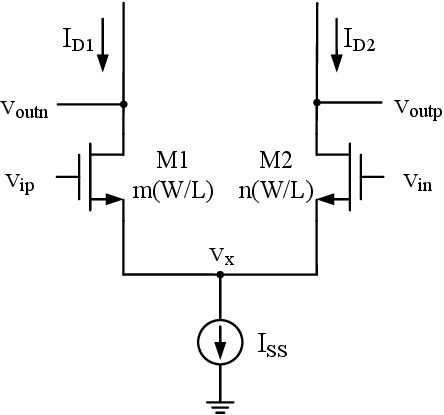}
\caption{Asymmetric differential pair.}
\label{fig_2}
\end{figure}

When the width-to-length ratios of the input transistors M1 and M2 are unequal, as shown in Figure \ref{fig_2}, with the width-to-length ratio of M1 being $m(W/L)$ and that of M2 being $n(W/L)$, the transconductance of the asymmetric differential pair $G_\mathrm{m}$ is given by:
\begin{equation}
\label{deqn_ex1}
G_m=\frac{\partial{\Delta I}_\mathrm{D}}{\partial{\Delta V}_\mathrm{in}}=\frac{1}{\left(n+m\right)^2}\left(P+4mn\sqrt{\frac{C_\mathrm{ox}\mu_\mathrm{n}\frac{W}{L}}{2}}Q\right),
\end{equation}

\noindent where
\begin{equation}
\label{deqn_ex2}
P=-2C_\mathrm{ox}\mu_\mathrm{n}\frac{W}{L}mn\left(m-n\right){\Delta V}_\mathrm{in},
\end{equation}
\begin{equation}
\label{deqn_ex3}
Q=\frac{\left(n+m\right)I_\mathrm{{SS}}-C_\mathrm{ox}\mu_\mathrm{n}\frac{W}{L}mn{{\Delta V}_\mathrm{in}}^2}{\sqrt{\left(n+m\right)I_\mathrm{SS}-\frac{C_\mathrm{ox}\mu_\mathrm{n}\frac{W}{L}}{2}mn{{\Delta V}_\mathrm{in}}^2}}.
\end{equation}

In equation (\ref{deqn_ex1}), $\mu_\mathrm{n}$ represents the mobility of the electron, $C_\mathrm{ox}$ is the unit capacitance of the gate oxide layer, $I_\mathrm{SS}$ is the tail current, and $\Delta I_\mathrm{D}$ is the differential output current.

\begin{figure}[htb]
\centering
\includegraphics[width=3.3in]{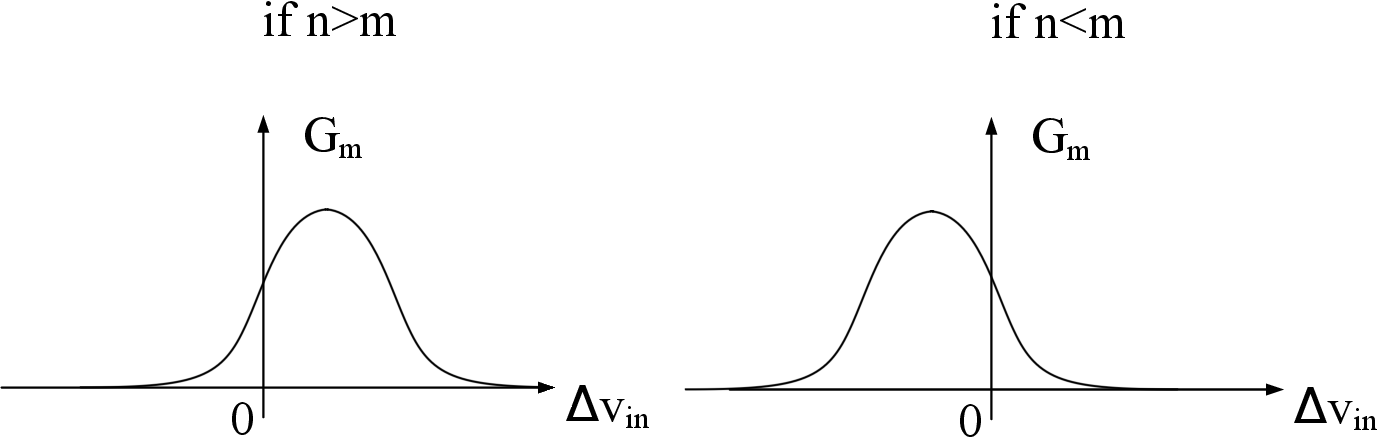}
\caption{$G_\mathrm{m}-\Delta V_\mathrm{in}$ curve of asymmetric differential pair.}
\label{fig_3}
\end{figure}

From equation (\ref{deqn_ex1}), the corresponding $G_\mathrm{m}-\Delta V_\mathrm{in}$ curve can be drawn. As shown in Figure \ref{fig_3}, if $n>m$, the $G_\mathrm{m}-\Delta V_\mathrm{in}$ curve shifts to the positive direction. Similarly, if $n<m$, the $G_\mathrm{m}-\Delta V_\mathrm{in}$ curve shifts to the negative direction.

\begin{figure}[htb]
\centering
\includegraphics[width=3in]{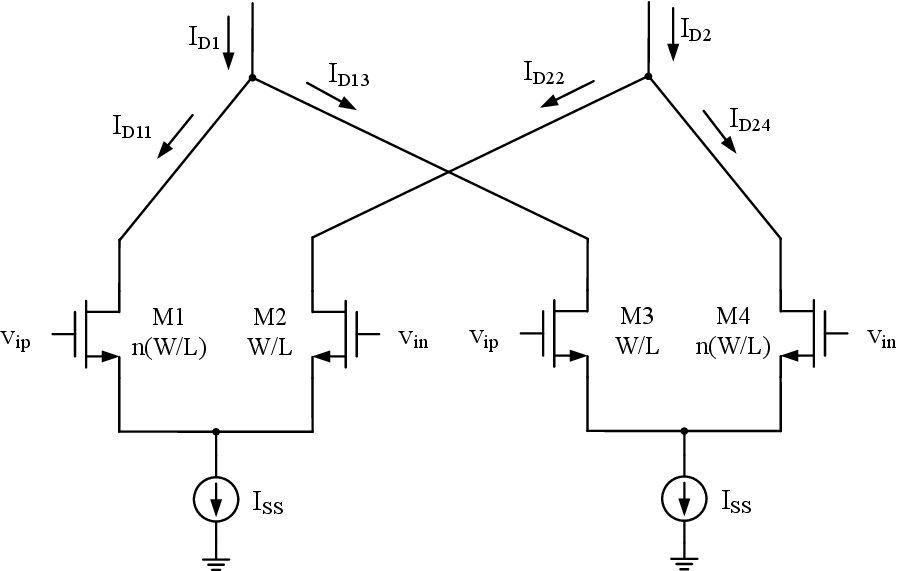}
\caption{Composite differential pair.}
\label{fig_4}
\end{figure}

As shown in Figure \ref{fig_4}, when two asymmetric differential pairs which are mirror images of each other are combined, they form a composite differential pair. The total transconductance of the composite differential pair $G_\mathrm{m}$ is equal to the sum of the transconductance of the two mirrored differential pairs.

\begin{figure}[htb]
\centering
\includegraphics[width=2.5in]{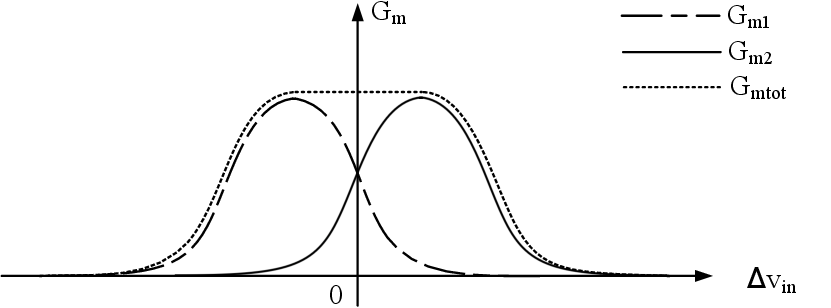}
\caption{$G_\mathrm{m}-\Delta V_\mathrm{in}$ curve of the composite differential pair.}
\label{fig_5}
\end{figure}

The $G_\mathrm{m}-\Delta V_\mathrm{in}$ curve of the composite differential pair is shown in Figure \ref{fig_5}. By appropriately choosing the width-to-length ratios of the transistors in each asymmetric differential pair, the effective transconductance of the composite differential pair can remain nearly constant within a certain input voltage range.

\section{The Structure of the Proposed Amplifier}
Based on the principle of transconductance linearization discussed in Section II, the main part of the proposed amplifier, as shown in Figure 6, employs a the composite differential pair, containing two asymmetric differential pairs, thereby improving the amplifier’s linearity under large input swing conditions.
\begin{figure}[htb]
\centering
\includegraphics[width=3.3in]{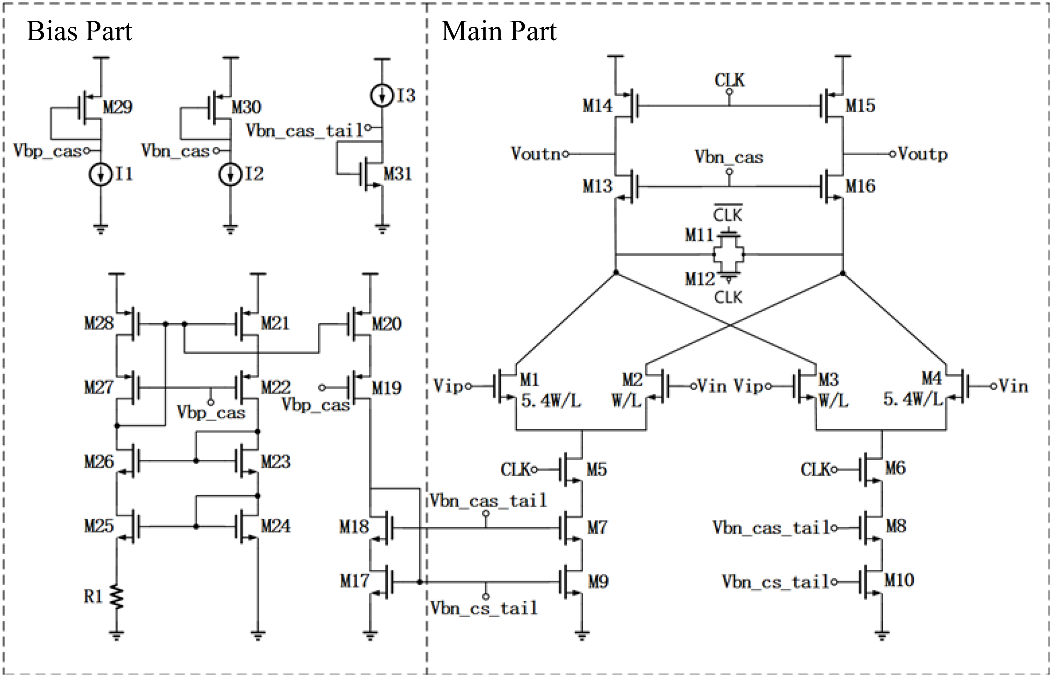}
\caption{The circuit of the proposed amplifier.}
\label{fig_6}
\end{figure}

The bias part of the proposed amplifier employs a constant transconductance bias circuit \cite{ref6}. The transconductance of M24 can be expressed as:
\begin{equation}
\label{deqn_ex4}
\left(g_\mathrm{m}\right)_\mathrm{M24}=\frac{2\left(1-\sqrt{\frac{\left(\frac{W}{L}\right)_\mathrm{M24}}{\left(\frac{W}{L}\right)_\mathrm{M25}}}\right)}{R}.
\end{equation}

In equation (\ref{deqn_ex4}), $\left(\frac{W}{L}\right)_\mathrm{M24}$ and  $\left(\frac{W}{L}\right)_\mathrm{M25}$ are the width-to-length ratio of M24 and M25, and $R$ is the resistance of resistor R1. It can be seen that $\left(g_m\right)_\mathrm{M24}$ depends only on $\frac{(W/L)_\mathrm{M24}}{(W/L)_\mathrm{M25}}$ and $R$ [7], and is not influenced by temperature or power supply voltage. Thus, the transconductance of the transistors in the bias circuit is relatively insensitive to variations in temperature and supply voltage. By copying currents from the bias circuit through the current mirror, the transconductance and the gain of the amplifier’s main part can exhibit high stability of temperature and power supply voltage. Let the current through M24 and M25 be $I$, the current amplification factor of the current mirror be $N$, so the current through M9 and M10 is $NI$. If the width-to-length ratio of M1 is $\left(\frac{W}{L}\right)_\mathrm{M1}$, $NmI$ of the tail current flows into one of the transistors in the differential pair, where $m$ depends on the width-to-length ratio of the transistors in the differential pair. For M1, $m=\frac{5.4}{6.4}$. Thus, the transconductance of M1 can be obtained as:
\begin{equation}
\label{deqn_ex5}
\left(g_\mathrm{m}\right)_\mathrm{M1}=2\sqrt{Nm}\frac{\sqrt{\frac{\left(\frac{W}{L}\right)_\mathrm{M1}}{\left(\frac{W}{L}\right)_\mathrm{M24}}}\left(1-\sqrt{\frac{\left(\frac{W}{L}\right)_\mathrm{M24}}{\left(\frac{W}{L}\right)_\mathrm{M25}}}\right)}{R}.
\end{equation}

From equation (\ref{deqn_ex5}), it can be seen that the transconductance of M1 depends only on $N$, $m$, $\left(\frac{W}{L}\right)_\mathrm{M1}$, $\left(\frac{W}{L}\right)_\mathrm{M24}$, $\left(\frac{W}{L}\right)_\mathrm{M25}$, and $R$. Since these values are fixed, the transconductance of M1 is not influenced by temperature or power supply voltage. Similarly, the transconductance of transistors M2, M3, and M4 is also constant. For a Gm-C dynamic amplifier, the gain is given by $\frac{G_\mathrm{m}T}{C}$, therefore the capacitor $C$ can be implemented using Metal-Insulator-Metal (MIM) or Metal-Oxide-Metal (MOM) capacitors \cite{ref1}, \cite{ref2}. The time constant $T$ can be provided by a divided signal of a external clock or the output signal from a DLL. They can ensure that the gain of the Gm-C dynamic amplifier maintains a high stability of temperature and power supply voltage.

Additionally, common-gate transistors M7, M8, M18, M19, M22, M23, M26, and M27 are used to isolate the drain voltage of the current mirror’s common-source transistors from its output voltage \cite{ref10}, equalizing the drain voltages of the current mirror’s input and output branches and improving current replication accuracy. Transistors M11, M12, M14, and M15 can reset the amplifier and reduce residual charge effects between phases. M13 and M16 suppress the impact of $V_\mathrm{outp}$ and $V_\mathrm{outn}$ on the drain currents of M1–M4. M7 and M8 also increase the tail current source’s output impedance, minimizing the effect of the input common-mode voltage on the tail current and thus on the differential pair’s transconductance. M5 and M6 control the amplifier’s operation between amplification and reset phases.

When the proposed amplifier is used as the residue amplifier in a pipelined SAR ADC, its gain is still influenced by process corners. Therefore, after the ADC is manufactured, the gain of the amplifier must be calibrated once by adjusting the bias circuit's resistor $R_1$ or through other methods.

\section{The Circuit Layout and The Simulation Results}
The circuit layout of the proposed amplifier is shown in Figure 7. The total area of the chip is 34,258.6192 ${\mu \mathrm{m}}^2$ (236.56 $\mu \mathrm{m}$×144.82 $\mu \mathrm{m}$).
\begin{figure}[htb]
\centering
\includegraphics[width=3.1in]{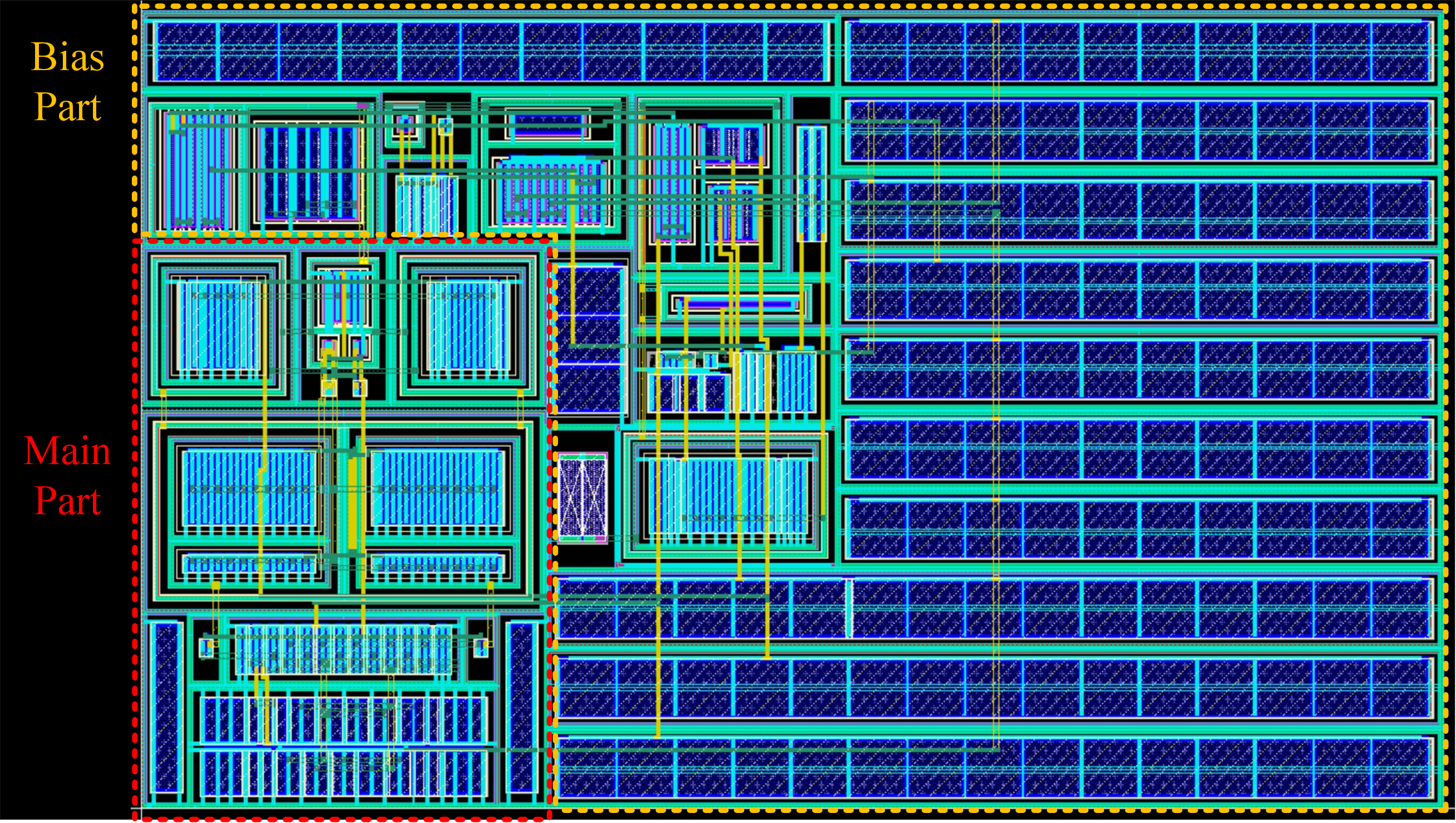}
\caption{The circuit layout of the proposed amplifier.}
\label{fig_7}
\end{figure}

\begin{figure}[htb]
\centering
\includegraphics[width=3.1in]{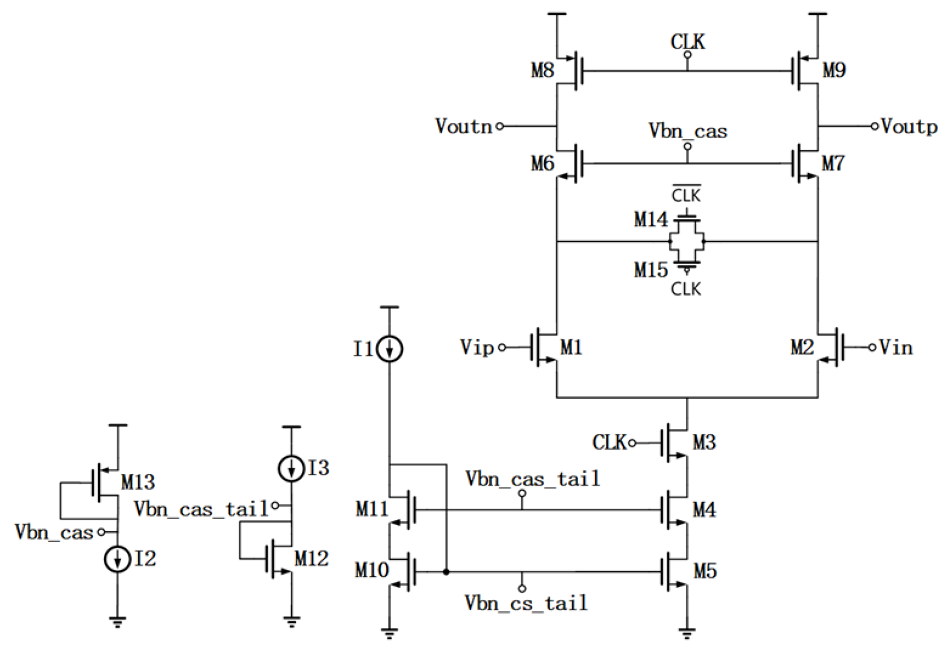}
\caption{The structure of the traditional Gm-C dynamic amplifier.}
\label{fig_8}
\end{figure}

The advantages of the proposed amplifier are demonstrated through simulation results in comparison with a traditional Gm-C dynamic amplifier. The structure of the traditional Gm-C dynamic amplifier is shown in Figure \ref{fig_8}.

\begin{figure}[htb]
\centering
\includegraphics[width=3.1in]{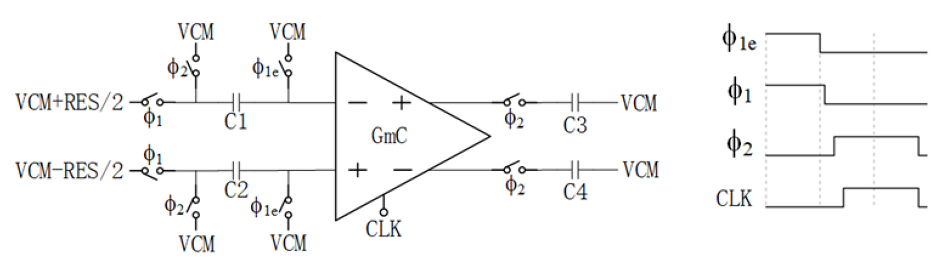}
\caption{The schematic of the simulation environment and the timing diagram of the control switches.}
\label{fig_9}
\end{figure}

The schematic of the simulation configuration and the timing diagram of the control switches are shown in Figure \ref{fig_9}. This configuration simulates the application in which a Gm-C dynamic amplifier is utilized as a residue amplifier in a pipelined SAR ADC. In simulations, the differential output voltage of the amplifier is measured at the falling edge of $\phi_\mathrm{2}$.

Simulations are conducted using the TSMC 180nm CMOS process with the process corner tt. In simulation A and B, the temperature is set to 27$\mathrm{^\circ C}$, and the power supply voltage is set to 5 V. The simulation results are as follows. 

\subsection{Transfer Characteristics and Gain versus Input Relationship}
The signal RES is configured as different DC voltage signals. By sweeping the DC value of RES and recording the corresponding differential output voltage, the transfer characteristics can be drawn, and the gain is calculated as the ratio of the differential output voltage to the value of RES.

\begin{figure}[htb]
\centering
\includegraphics[width=3.3in]{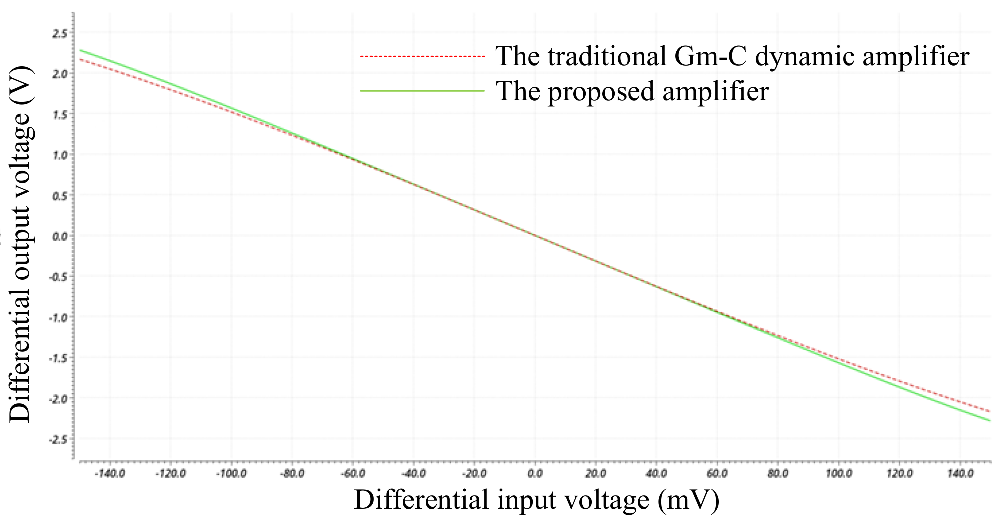}
\caption{Input–output transfer characteristics.}
\label{fig_10}
\end{figure}

Figure \ref{fig_10} shows the input–output transfer characteristics of the two amplifiers. As the absolute value of the differential input voltage exceeds 60 mV, the transfer curve of the traditional Gm-C dynamic amplifier begins to bend, indicating nonlinearity and a noticeable drop in gain. In contrast, the proposed amplifier maintains a more linear transfer characteristic within the same input range.

\begin{figure}[htb]
\centering
\includegraphics[width=3.3in]{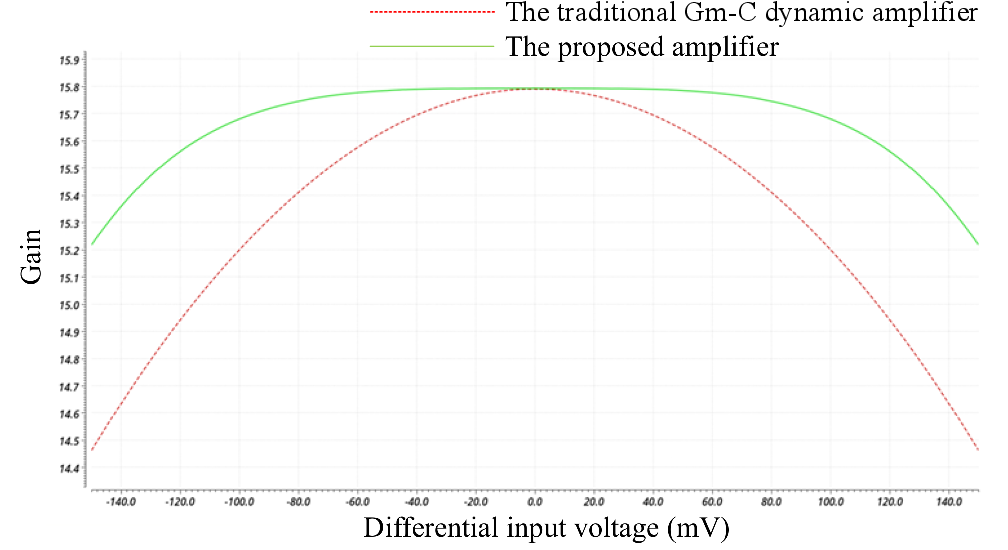}
\caption{Gain versus input relationship.}
\label{fig_11}
\end{figure}

Figure \ref{fig_11} shows the gain versus input relationship of the two amplifiers. For the traditional Gm-C dynamic amplifier, the gain peaks at a differential input of 0 V and decreases sharply as the magnitude of the differential input voltage increases, exhibiting significant nonlinearity. In contrast, the proposed amplifier maintains a relatively constant gain across the input range of -40 mV to +40 mV, demonstrating improved linearity.

Taken together, Figures \ref{fig_10} and \ref{fig_11} clearly demonstrate that the proposed amplifier exhibits significantly better linearity than the traditional Gm-C dynamic amplifier.

\subsection{Spectral Characteristics of the Output Signal and THD}
The signal RES is configured as a sinusoidal signal with an amplitude of 60 mV and a frequency of $\frac{3\times2\ }{1024}$ MHz. The clock frequency is 2 MHz. The amplifier is triggered to perform 1024 amplification cycles, yielding 1024 output samples. A Fast Fourier Transform (FFT) is then performed on these output samples to obtain the frequency spectrum of the amplifier’s output signal.

\begin{figure}[htb]
\centering
\includegraphics[width=3.3in]{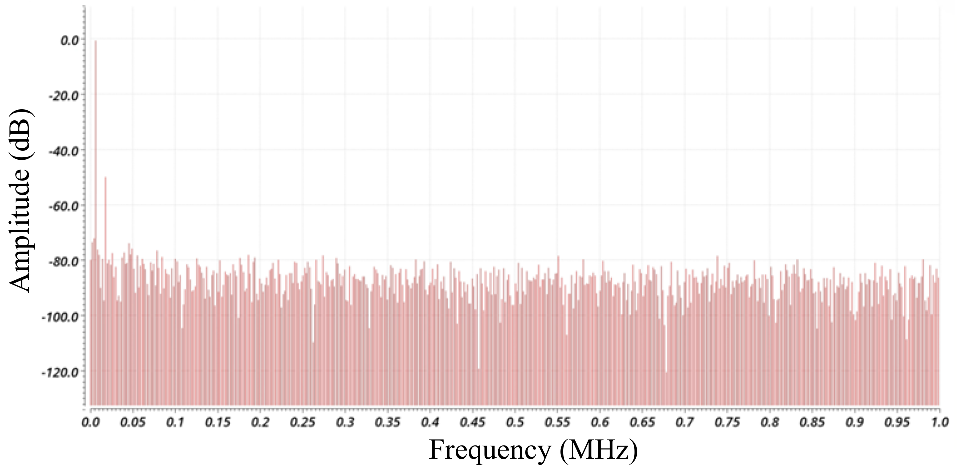}
\caption{Frequency spectrum of the sampled output signal from the traditional Gm-C dynamic amplifier.}
\label{fig_12}
\end{figure}

Figure \ref{fig_12} shows the frequency spectrum of the sampled output signal from the traditional Gm-C dynamic amplifier. In addition to the dominant component corresponding to the input signal, a significant harmonic component is observed. Its total harmonic distortion (THD) is 49.3 dB, indicating that the input signal undergoes considerable distortion after amplification by the traditional Gm-C dynamic amplifier.

\begin{figure}[htb]
\centering
\includegraphics[width=3.3in]{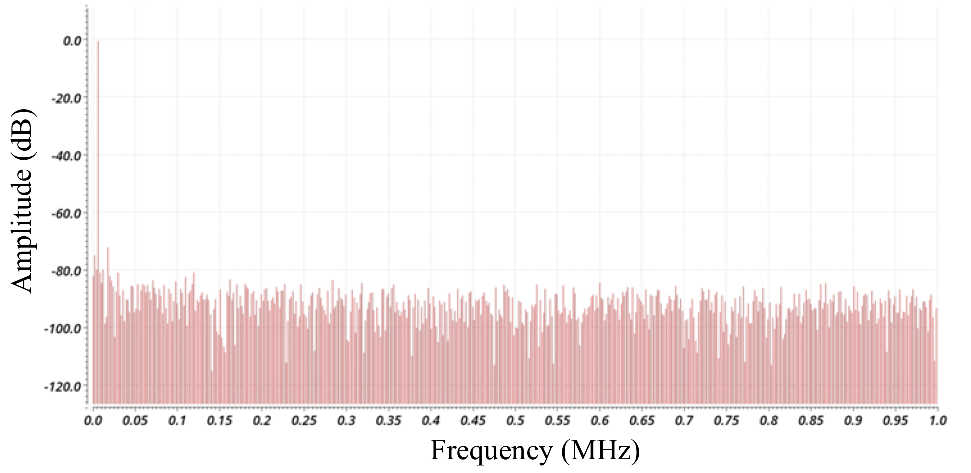}
\caption{Frequency spectrum of the sampled output from the proposed amplifier.}
\label{fig_13}
\end{figure}

In contrast, Figure \ref{fig_13} presents the frequency spectrum of the sampled output signal from the proposed amplifier. Its THD is 70.5 dB, which is 21.2 dB lower than that of the traditional Gm-C dynamic amplifier, demonstrating a substantially reduced level of distortion and a improved linearity.

\subsection{Sensitivity to Temperature and Power Supply Voltage Variations}

In this simulation, the signal RES is configured as a 1 mV DC voltage signal. Temperature is swept from -40 $\mathrm{^\circ C}$ to 120 $\mathrm{^\circ C}$, and the power supply voltage is varied by $\pm$10\% around its nominal value of 5 V, i.e. from 4.5 V to 5.5 V. The gain is calculated as the ratio of the differential output voltage to the value of RES.

\begin{figure}[htb]
\centering
\includegraphics[width=3.3in]{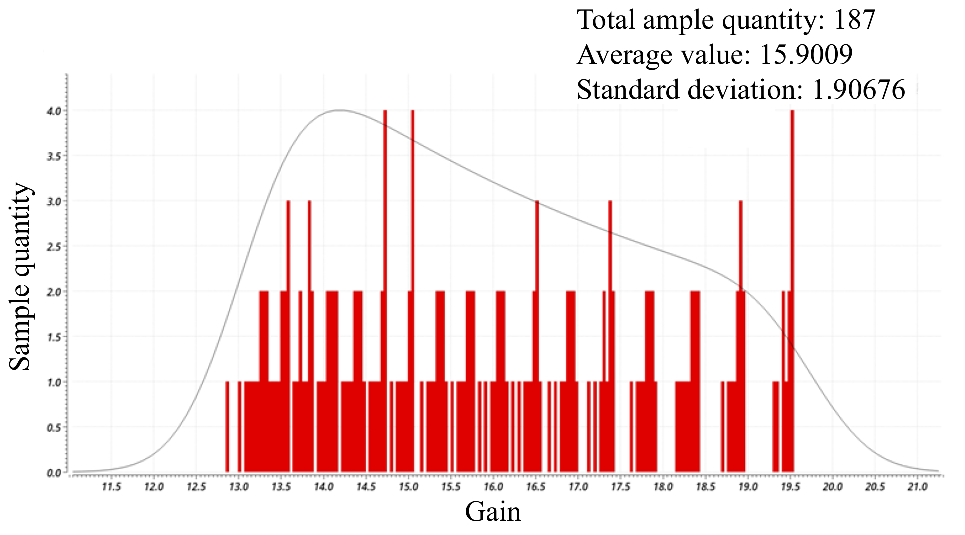}
\caption{Gain distribution of the traditional Gm-C dynamic amplifier.}
\label{fig_14}
\end{figure}

Figure \ref{fig_14} presents the gain distribution of the traditional Gm-C dynamic amplifier under varying temperature and power supply voltage conditions. The standard deviation of the gain is 1.9, and the distribution range of the gain is from 13 to 19.5.

\begin{figure}[htb]
\centering
\includegraphics[width=3.3in]{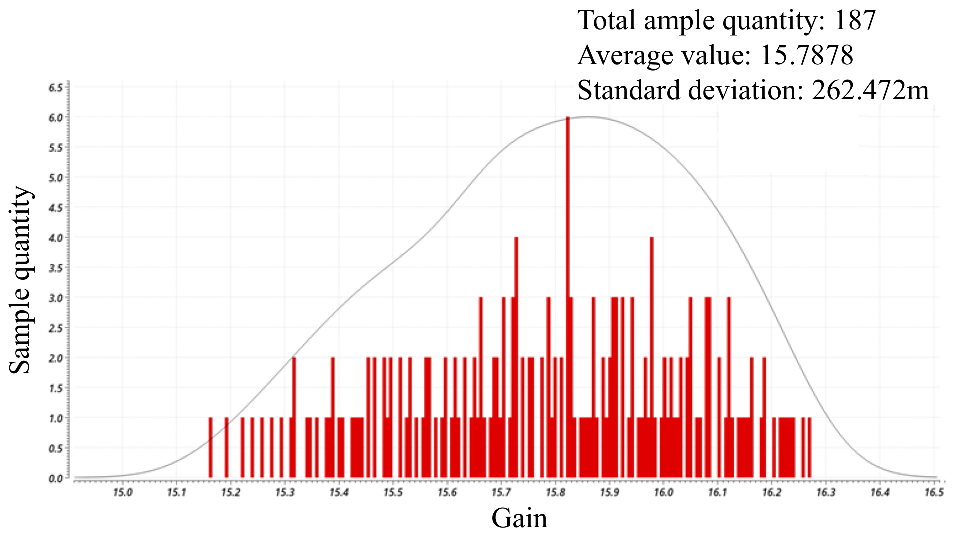}
\caption{Gain distribution of the proposed amplifier.}
\label{fig_15}
\end{figure}

In comparison, Figure \ref{fig_15} shows the gain distribution of the proposed amplifier. The standard deviation of the gain is 262m, which is comparable to the gain variation reported in \cite{ref8}, and the distribution range of the gain is from 15.1 to 16.3.

These results indicate that, compared to the traditional Gm-C dynamic amplifier, the proposed amplifier exhibits significantly lower gain variation under temperature and power supply voltage fluctuations, demonstrating higher temperature and power supply voltage stability.

\section{Conclusion}
This paper presents a Gm-C dynamic amplifier with high linearity and high temperature and power supply voltage stability. Two asymmetric differential pairs and a constant-gm bias circuit are utilized to enhance its stability and linearity. The layout of the chip is drawn and the total area of the chip is 34,258.6192 ${\mu \mathrm{m}}^2$. Simulations based on the TSMC 180nm CMOS process show that the proposed amplifier maintains a relatively constant gain over a differential input range of -40 mV to +40 mV, and its THD is 70.5 dB. When the differential input is 1 mV, the power supply voltage varies by $\pm$10\% and the temperature ranges from -40 $\mathrm{^\circ C}$ to 120 $\mathrm{^\circ C}$, the standard deviation of the gain is 262m, and the distribution of the gain is from 15.1 to 16.3. The proposed amplifier preforms better than the traditional Gm-C dynamic amplifier and is well-suited for use as the residue amplifier in high-speed, high-precision, low-power pipelined SAR ADCs.

\end{document}